\documentclass[aps,prd,nofootinbib,amsmath,amssymb,superscriptaddress,twocolumn,10pt]{revtex4}
\let\sb=_ \catcode`\_=\active \def_#1{\ensuremath \sb{\rm#1}}
\usepackage{amsmath}
\usepackage{amsfonts}
\usepackage{txfonts}
\usepackage{graphicx}
\usepackage{dcolumn}
\usepackage{bbm}
\usepackage{amssymb}
\usepackage{latexsym}
\usepackage{CJK}
\usepackage{titlesec}
\usepackage{color}
\usepackage[colorlinks=true, linkcolor=red, citecolor=blue]{hyperref}
\begin{document}
 \bibliographystyle{unsrt}

\title{Diagnosing holographic dark energy models with statefinder hierarchy}
\author{Jing-Fei Zhang}
\affiliation{Department of Physics, College of Sciences, Northeastern University,
Shenyang 110004, China}
\author{Jing-Lei Cui}
\affiliation{Department of Physics, College of Sciences, Northeastern University,
Shenyang 110004, China}
\author{Xin Zhang\footnote{Corresponding author}}
\email{zhangxin@mail.neu.edu.cn} \affiliation{Department of Physics, College of Sciences,
Northeastern University, Shenyang 110004, China}
\affiliation{Center for High Energy Physics, Peking University, Beijing 100080, China}

\begin{abstract}
 We apply a series of null diagnostics based on the statefinder hierarchy to diagnose different holographic dark energy models including the original holographic dark energy, the new holographic dark energy, the new agegraphic dark energy, and the Ricci dark energy models. We plot the curves of statefinders $S^{(1)}_3$ and $S^{(1)}_4$ versus redshift $z$ and the evolutionary trajectories of $\{S^{(1)}_3, \epsilon\}$ and $\{S^{(1)}_4, \epsilon\}$ for these models, where $\epsilon$ is the fractional growth parameter. Combining the evolution curves with the current values of $S^{(1)}_3$, $S^{(1)}_4$, and $\epsilon$, we find that the statefinder $S^{(1)}_4$ performs better than $S^{(1)}_3$ for diagnosing the holographic dark energy models. In addition, the conjunction of the statefinder hierarchy and the fractional growth parameter is proven to be a useful method to diagnose the holographic dark energy models, especially for breaking the degeneracy of the new agegraphic dark energy model with different parameter values.
\end{abstract}
\pacs{95.36.+x, 98.80.Es, 98.80.-k}
\maketitle

\renewcommand{\thesection}{\arabic{section}}
\renewcommand{\thesubsection}{\arabic{subsection}}
\titleformat*{\section}{\flushleft\bf}
\titleformat*{\subsection}{\flushleft\bf}
\section {Introduction}
Dark energy (DE) with negative pressure was considered as an exotic component causing the Universe to a stage of accelerating expansion and has been widely studied~\cite{lamada}. Because of the lack of knowledge about the nature of DE, physicists constructed a host of viable theoretical DE models. The $\Lambda$CDM model consisting of the cosmological constant ($\Lambda$) and the cold dark matter (CDM) is the simplest one, in which DE has the equation of state $w=-1$. And this elegant model is even defined as a criterion in several cosmological observations. However, the cosmological constant scenario has to face the so-called ``fine-tuning problem" and ``coincidence problem". Furthermore, different observational data are in tension with one another to some extent when constraining parameters of the $\Lambda$CDM model. Under such circumstances, the possibility that $w$ is dependent on time cannot be excluded. At the present, a number of dynamical DE models have been suggested, such as quintessence~\cite{quintessence}, quintom~\cite{Feng:2004ad}, k-essence~\cite{ArmendarizPicon:1999rj}, Chaplygin gas~\cite{Kamenshchik:2001cp}, and so on.

In face of numerous DE models, it is important to discriminate various models. Sahni et al.~\cite{Sahni:2002fz} introduced the statefinder diagnostic $\{r,~s\}$, which is a geometrical diagnosis in a model-independent manner. The statefinder parameter pair $\{r,~s\}$ contains the third-derivative of $a(t)$, where $a(t)$ is the scale factor of the Universe. Since different DE models exhibit different evolution trajectories in the $r$--$s$ plane, and especially can be separated distinctively with the values of $\{r_0,~s_0\}$, the statefinder can be used to diagnose different DE models~\cite{sfide}. Besides, other diagnostics, such as $Om$ and $Om3$~\cite{om,Shafieloo:2012rs,Sahni:2014ooa}, were also used to distinguish the DE models. In the previous work~\cite{Cui:2014sma}, we compared the holographic DE models by using the statefinder pair $\{r,~s\}$. Here, the holographic DE models include the original holographic dark energy (HDE)~\cite{Li:2004rb}, the new holographic dark energy (NHDE)~\cite{Li:2012xf}, the new agegraphic dark energy (NADE)~\cite{Wei:2007ty}, and the Ricci dark energy (RDE)~\cite{Gao:2007ep}, which were all proposed based on the holographic principle. All these holographic DE models can be used to interpret and describe the cosmic acceleration~\cite{hde1,hde2}. By employing the statefinder diagnostic the holographic DE models can be differentiated effectively in the low-redshift region~\cite{Cui:2014sma}. Also the holographic DE models with different parameter values can be distinguished, except for the NADE model. The $r(z)$ and $r(s)$ curves of NADE with different parameter values are in strong degeneracy during the whole evolution history~\cite{Cui:2014sma}.

To break the degeneracy, we will take into account the statefinder hierarchy~\cite{Arabsalmani:2011fz} with higher derivatives of $a(t)$. In Ref.~\cite{Arabsalmani:2011fz}, the original aim of introducing the statefinder hierarchy is to distinguish $\Lambda$CDM model from evolving models and to extend the null diagnostic.
The statefinder hierarchy has been applied to study some dynamical DE and modified gravity models; see, e.g., Refs.~\cite{sfh1,sfh2}.
In this paper, we will use this diagnostic to discriminate deeply similar dynamical DE models such as the holographic type DE models. Today's values of statefinder parameters which can be extracted theoretically from the low-redshift observational data are helpful to differentiate DE models. In this paper, the current values of the statefinders are also used to differentiate the holographic DE models, and to break the degeneracy mentioned above. Furthermore, the fractional growth parameter $\epsilon (z)$ is also proposed as a supplement to null diagnostic~\cite{Arabsalmani:2011fz}. Thus, we also combine the statefinder hierarchy with the fractional growth parameter to differentiate the holographic DE models, following the method proposed in Ref.~\cite{Arabsalmani:2011fz}.

In this paper, we use the statefinder hierarchy to diagnose holographic DE models including HDE, NHDE, NADE, and RDE. In Sects. 2 and 3, the diagnostic methods and a series of holographic DE models are briefly reviewed, respectively. Diagnosing holographic DE models with the statefinder hierarchy will be presented in Sect. 4. The conclusion is given in Sect. 5.

\section {The statefinder hierarchy and the growth rate of perturbations}
This section consists of two parts. In the first part, we introduce the general expressions of the statefinder hierarchy~\cite{Arabsalmani:2011fz}, and give the specific expressions of them which contain variables $\Omega_{de}$ and $w$ dependent on redshift $z$,  where $\Omega_{de}$ is the fractional density of DE ($\Omega_{de}\equiv \rho_{de}/{3M^2_p}H^2$) and $w$ is the equation of state (EOS) of DE ($w\equiv p_{de}/\rho_{de}$). The growth rate of perturbations is briefly described in the second part.
\subsection*{2.1 The statefinder hierarchy}
In this paper, we consider a spatially flat Friedmann-Robertson-Walker (FRW) universe containing dark energy and matter. The Friedmann equation is
 \begin{equation}
H^2=\frac{1 }{3M^2_p}(\rho_{de}+\rho_m),
\end {equation}
where $H=\dot{a}/a$ is the Hubble parameter (the dot denotes the derivative with respect to time $t$), $M^2_p = (8\pi G)^{-1}$ is the reduced Planck mass, $\rho_{de}$ and $\rho_m$ are the energy densities for dark energy and matter, respectively.

The scale factor of the Universe, $a(t)/a_0=(1+z)^{-1}$, can be Taylor expanded around the present epoch $t_0$ as follows:
\begin{equation}
\frac{a(t)}{a_0}=1+\sum\limits_{\emph{n}=1}^{\infty}\frac{A_{\emph{n}}(t_0)}{n!}[H_0(t-t_0)]^n,
\end {equation}
where
\begin{equation}
A_{\emph{n}}=\frac{a(t)^{(n)}}{a(t)H^n},~~n\in N,
\end {equation}
with $a(t)^{(n)}=d^na(t)/dt^n$. Various derivatives of $a(t)$ have been described historically by other quantities. $A_2$ is the negative value of the deceleration parameter $q$, and $A_3$ is the statefinder $r$~\cite{Sahni:2002fz,Chiba:1998tc} or the jerk $j$~\cite{Visser:2003vq}. In addition, $A_4$ and $A_5$ are the snap $s$ and the lerk $l$~\cite{Visser:2003vq,Dabrowski:2005fg}, respectively.
For the $\Lambda$CDM model, we can easily get:
\begin{align}
&A_{2}=1-\frac{3}{2}\Omega_{m},\\
&A_{3}=1,\\
&A_{4}=1-\frac{3^2}{2}\Omega_{m},\\
&A_{5}=1+3\Omega_{\rm m}+\frac{3^3}{2}\Omega_{m}^{2},~~\rm{etc.},
\end{align}
where $ \Omega_{m}\equiv \rho_{m}/{3M^2_p}H^2$ is the fractional density of matter. The statefinder hierarchy, $S_{\emph{n}}$, is defined as~\cite{Arabsalmani:2011fz}:
\begin{align}
&S_{2}=A_{2}+\frac{3}{2}\Omega_{\rm m},\\
&S_{3}=A_{3},\\
&S_{4}=A_{4}+\frac{3^2}{2}\Omega_{\rm m},\\
&S_{5}=A_{5}-3\Omega_{\rm m}-\frac{3^3}{2}\Omega_{\rm m}^{2},~~\rm{etc.}
\end{align}
The reason for this redefinition is to peg the statefinder at unity for $\Lambda$CDM during the cosmic expansion,
\begin{eqnarray}
S_{\emph{n}}|_{\Lambda \rm{CDM}}=1.
\end{eqnarray}
This equation defines a series of null diagnostics for $\Lambda$CDM when $n\geq3$. By using this diagnostic, we can distinguish easily the $\Lambda$CDM model from other DE models. Because of $\Omega_{m}=\frac{2}{3}(1+q)$ for $\Lambda$CDM, when $n\geq3$, statefinder hierarchy can be rewritten as:
\begin{align}
&S^{(1)}_{3}=A_{3},\\
&S^{(1)}_{4}=A_{4}+3(1+q)\\
&S^{(1)}_{5}=A_{5}-2(4+3q)(1+q),~~\rm{etc.},
\end{align}
where the superscript $(1)$ is to discriminate between $S^{(1)}_{\emph{n}}$ and $S_{\emph{n}}$. Obviously, $S^{(1)}_{\emph{n}}|_{\Lambda \rm{CDM}}=1$ for $\Lambda$CDM and $S^{(1)}_{3}$ is statefinder $r$~\cite{Sahni:2002fz,Chiba:1998tc}.
In this paper, we use the statefinders $S^{(1)}_{3}$ and $S^{(1)}_{4}$ to diagnose the holographic type DE models. We give the specific expressions of $S^{(1)}_{3}$ and $S^{(1)}_{4}$ using the variables $\Omega_{de}$ and $w$ dependent on redshift $z$:
\begin{equation}
S^{(1)}_{3}=1+\frac{9}{2}\Omega_{de}w(1+w)-\frac{3}{2}\Omega_{de}w',\label{eq16}
\end{equation}
\begin{equation} \begin{aligned}
S^{(1)}_{4}=&1-\frac{27}{2}w(w+1)(w+\frac{7}{6})\Omega_{de}-\frac{27}{4}w^2(w+1)\Omega_{de}^2\\
&+\frac{3}{2}\Omega_{de}\left(w'(\frac{13}{2}+9w+\frac{3}{2}w\Omega_{de})-w''\right),\label{eq17}
\end{aligned} \end{equation}
 where the prime denotes the derivative with respect to $x=\ln a$.

\begin{figure*}[htbp]
\centering
\includegraphics[scale=0.5]{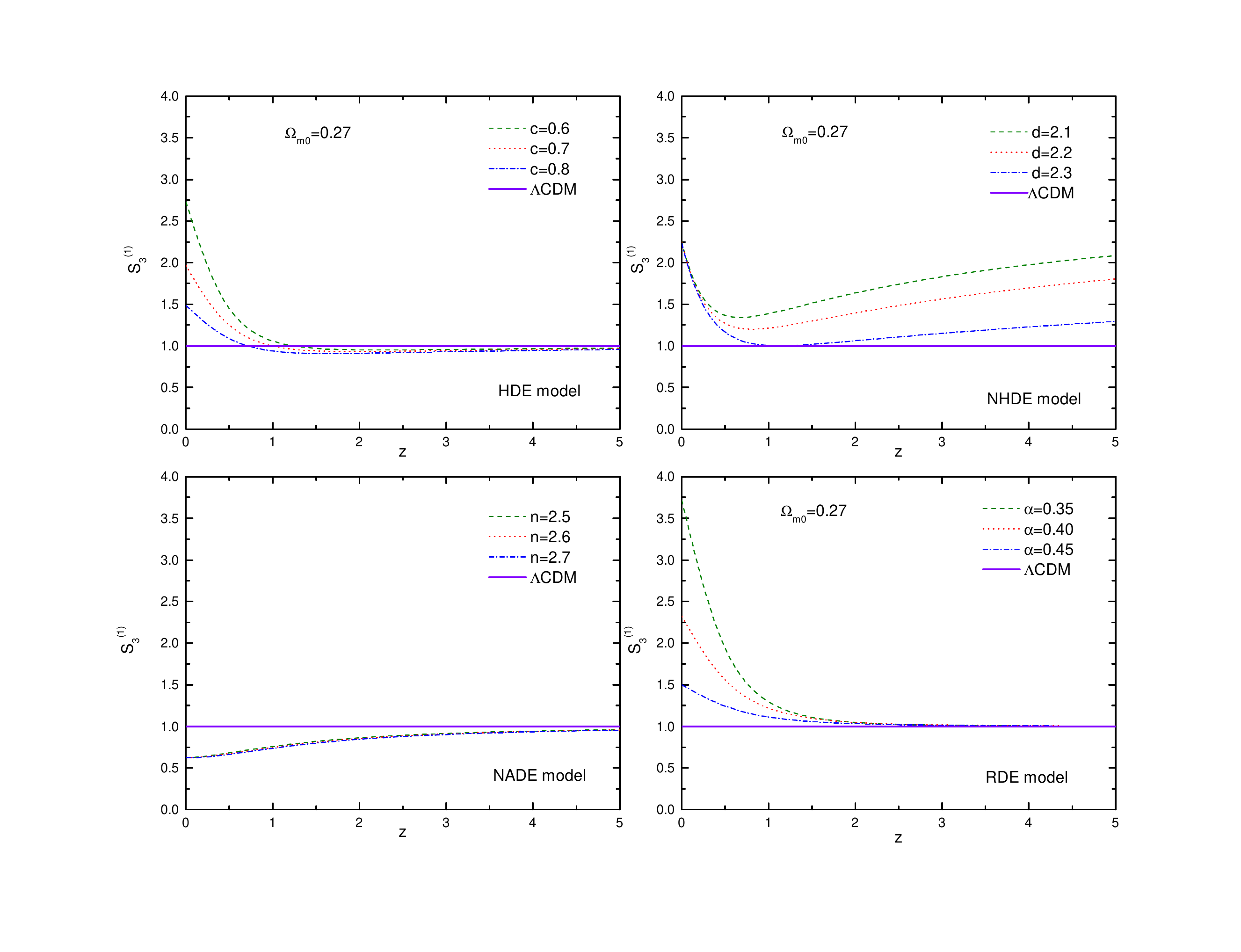}
\caption{\label{fig1} Evolutions of $S^{(1)}_3$ versus redshift $z$ for the HDE, NHDE, NADE, and RDE models. The $S^{(1)}_3$ curve of the $\Lambda$CDM model is also shown for comparison.}
\end{figure*}

\begin{figure*}[htbp]
\centering
\includegraphics[scale=0.5]{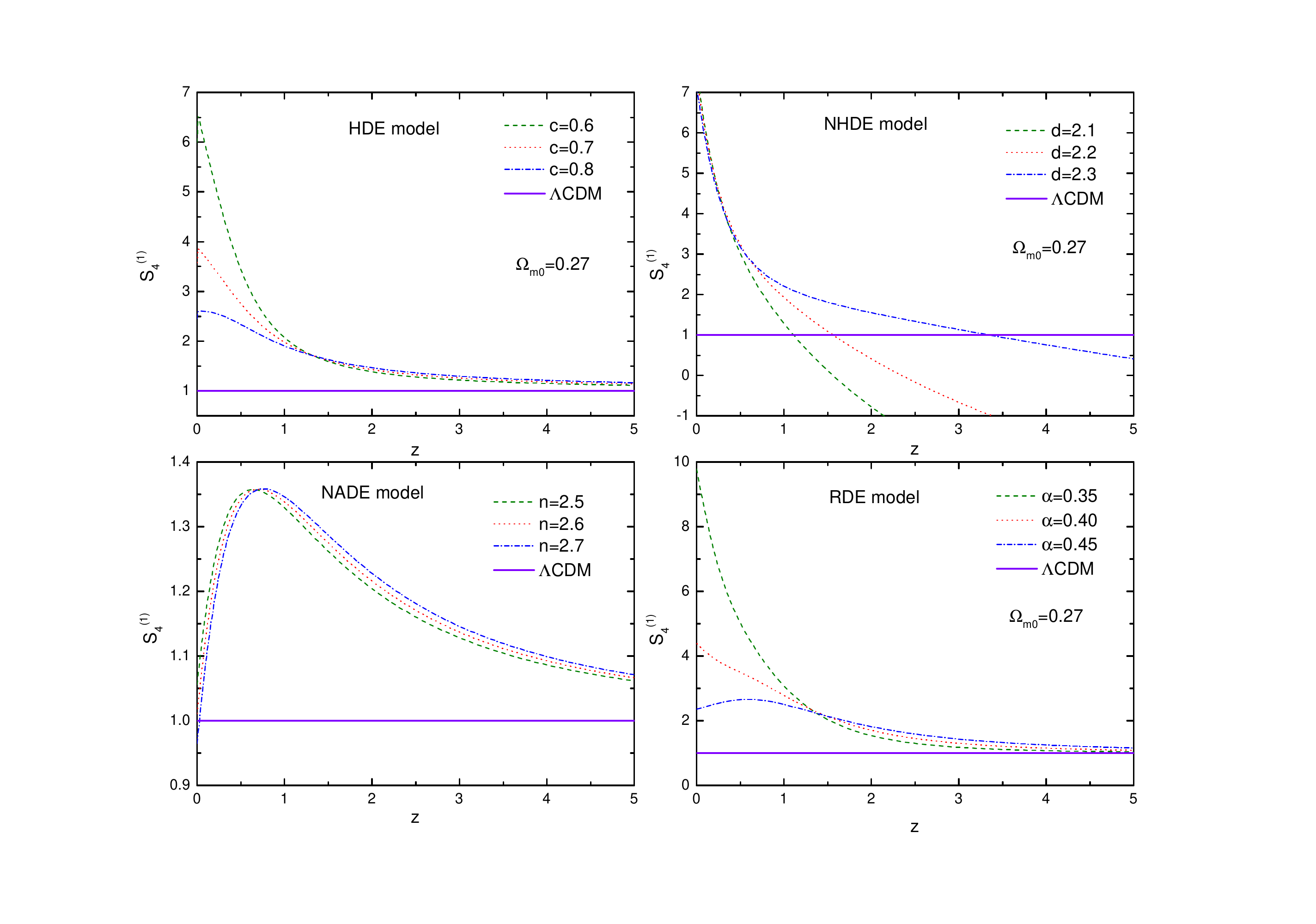}
\caption{\label{fig2} Evolutions of $S^{(1)}_4$ versus redshift $z$ for the HDE, NHDE, NADE, and RDE models. The $S^{(1)}_4$ curve of the $\Lambda$CDM model is also shown for comparison.}
\end{figure*}

\begin{table*}[tbp]
\centering
\begin{tabular}{|lcccccccccccccccc|}
\hline
 &\multicolumn{3}{c}{HDE}&&\multicolumn{3}{c}{NHDE}&&\multicolumn{3}{c}{NADE}&&\multicolumn{3}{c}{RDE}&\\
\cline{2-4}\cline{6-8}\cline{10-12}\cline{14-16}
Parameters&$c=0.6$&$c=0.7$&$c=0.8$&&$d=2.1$&$d=2.2$&$d=2.3$&&$n=2.5$&$n=2.6$&$n=2.7$&&$\alpha=0.35$&$\alpha=0.40$&$\alpha=0.45$&\\
\hline
$S^{(1)}_{3to}$&$2.73$&$1.97$&$1.49$&&$2.26$&$2.25$&$2.24$&&$0.62$&$0.62$&$0.62$&&$3.72$&$2.33$&$1.5$&\\
$S^{(1)}_{4to}$&$6.00$&$3.57$&$2.45$&&$163.58$&$177.43$&$191.82$&&$1.06$&$1.01$&$0.97$&&$9.78$&$4.40$&$2.36$&\\
$\epsilon _{0}$&$1.005$&$1.000$&$0.997$&&$0.994$&$0.996$&$0.999$&&$1.117$&$1.079$&$1.042$&&$1.009$&$1.002$&$0.996$&\\
\hline
$\Delta S^{(1)}_{3to}$&&$1.24$&&&&$0.02$&&&&$0$&&&&$2.22$&&\\
$\Delta S^{(1)}_{4to}$&&$3.55$&&&&$28.24$&&&&$0.09$&&&&$7.42$&&\\
$\Delta \epsilon _{0}$&&$0.008$&&&&$0.005$&&&&$0.075$&&&&$0.013$&&\\
\hline
\end{tabular}
\caption{\label{tab1} The present-day values of the statefinders and the fractional growth parameter, $S^{(1)}_{3to}$, $S^{(1)}_{4to}$, and $\epsilon _0$, and the differences of them, $\Delta S^{(1)}_{3to}$, $\Delta S^{(1)}_{4to}$, and $\Delta \epsilon _0$, for the holographic DE models, where $\Delta S^{(1)}_{3to}=S^{(1)}_{3to}(\rm{max})-S^{(1)}_{3to}(\rm{min})$, $\Delta S^{(1)}_{4to}=S^{(1)}_{4to}(\rm{max})-S^{(1)}_{4to}(\rm{min})$, and $\Delta \epsilon _{0}=\epsilon _0(\rm{max})-\epsilon _0(\rm{min})$ within one model.}
\end{table*}

\begin{figure}[htbp]
\centering
\includegraphics[scale=0.3]{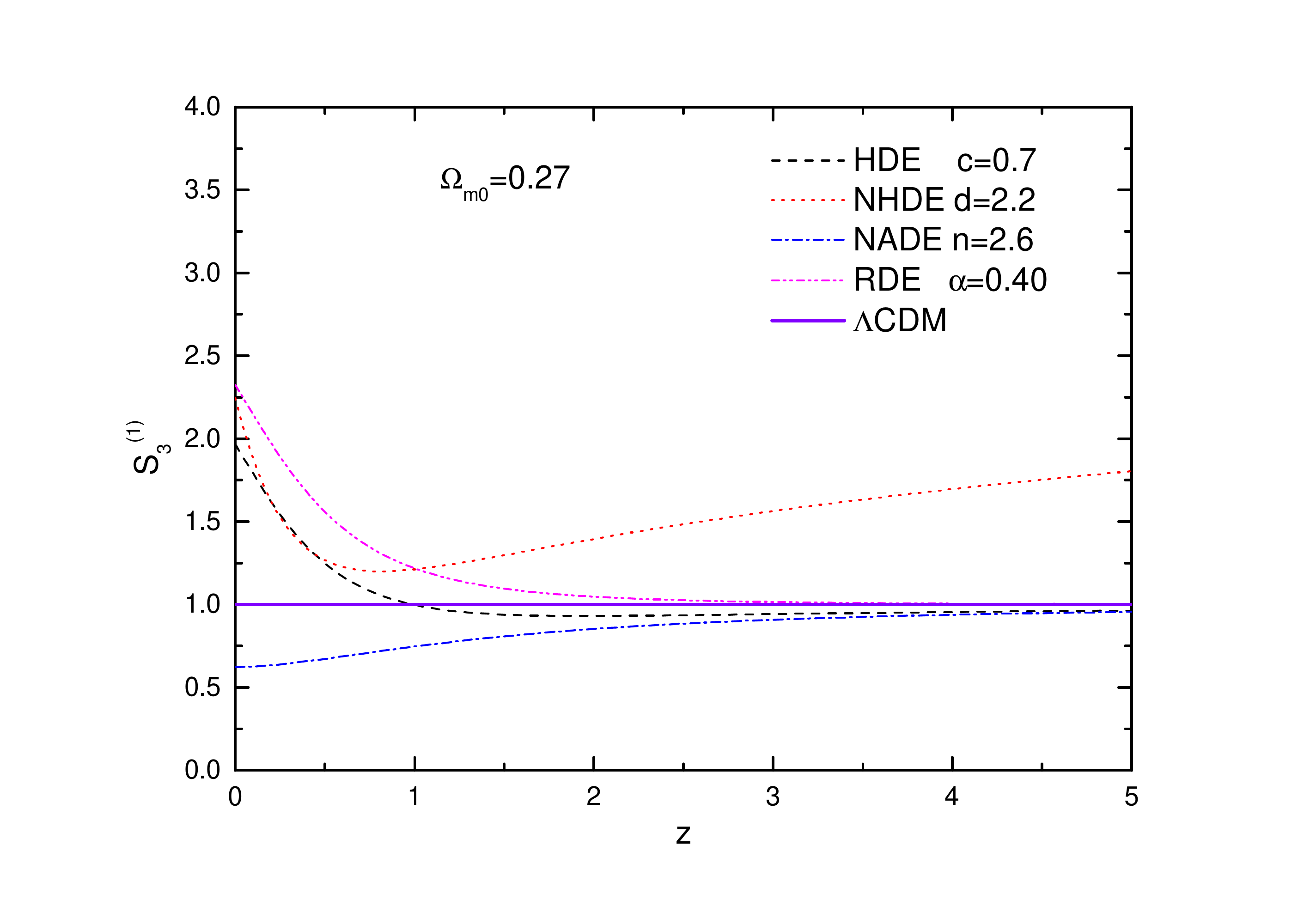}
\caption{\label{fig3} Comparisons of various holographic DE models in the $S^{(1)}_3$($z$) evolution diagram. The $\Lambda$CDM model is also shown for comparison.}
\end{figure}

\begin{figure}[htbp]
\centering
\includegraphics[scale=0.3]{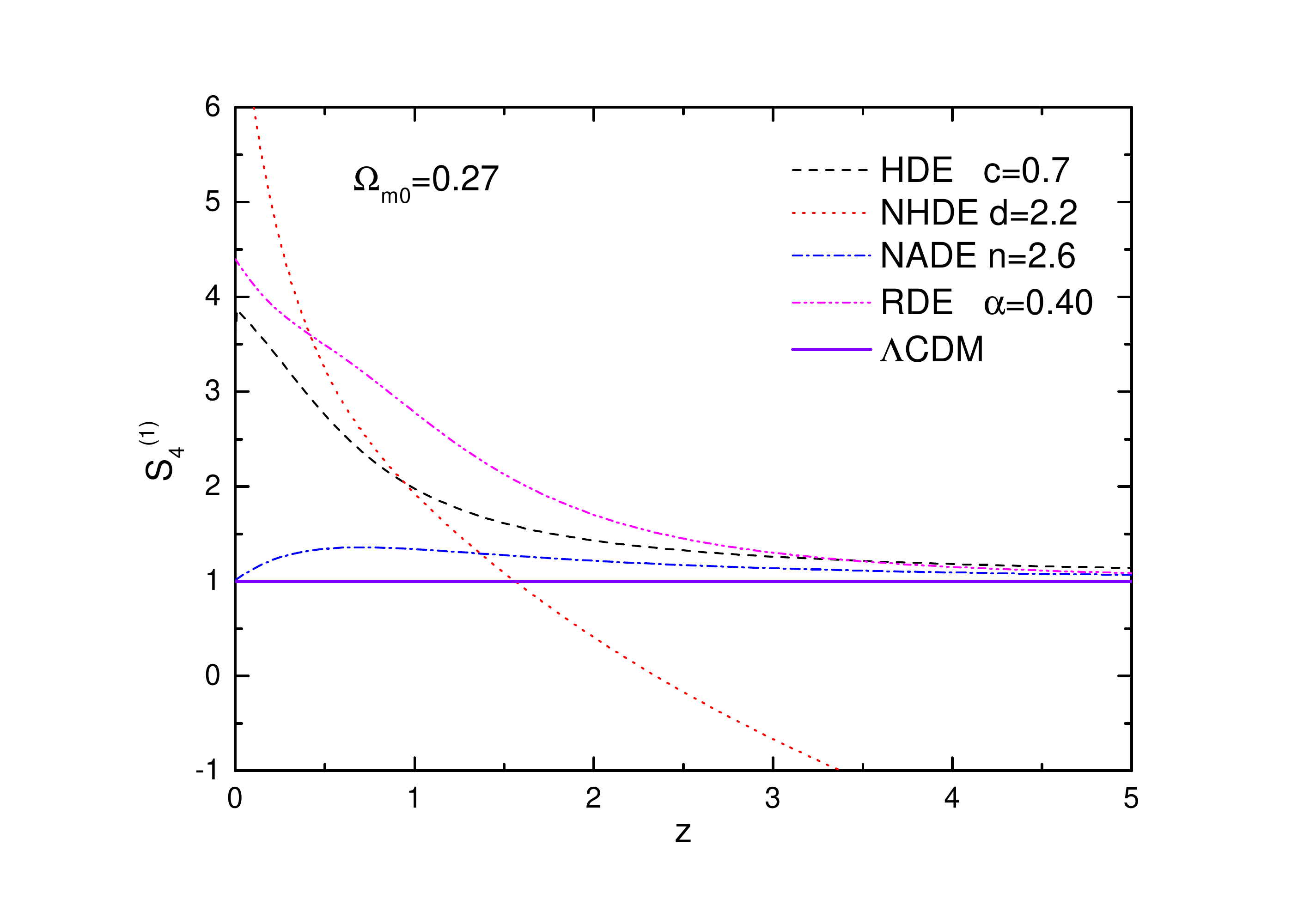}
\caption{\label{fig4} Comparisons of various holographic DE models in the $S^{(1)}_4$($z$) evolution diagram. The $\Lambda$CDM model is also shown for comparison.}
\end{figure}

\begin{figure*}[htbp]
\centering
\includegraphics[scale=0.5]{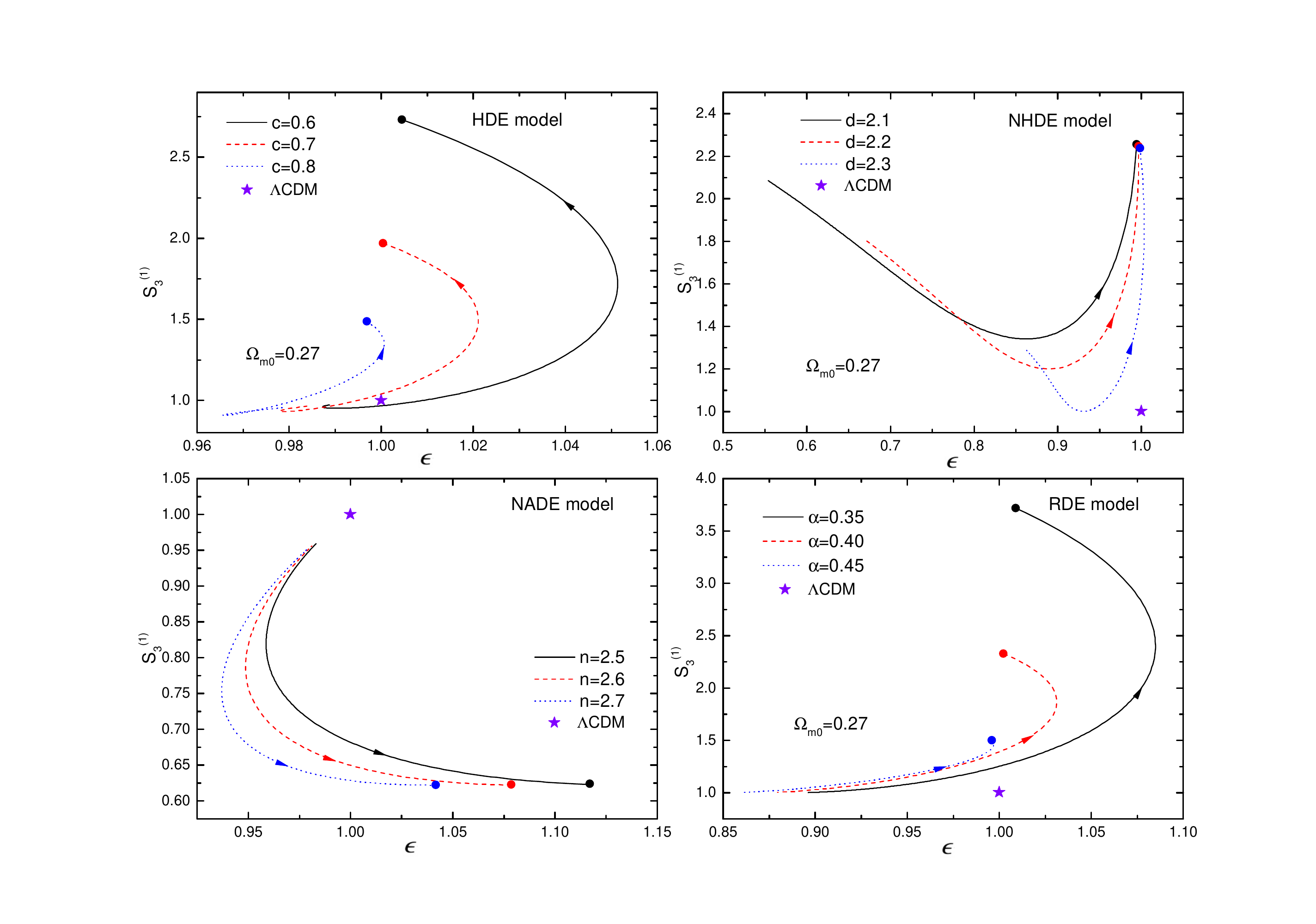}
\caption{\label{fig5} The composite null diagnostics $\{S^{(1)}_3, \epsilon\}$ are plotted for the HDE, NHDE, NADE and RDE models.
The current values of $\{S^{(1)}_3, \epsilon\}$ of the holographic DE models are marked by the round dots.
$\{S^{(1)}_3, \epsilon\}=\{1,1\}$ for the $\Lambda$CDM model is also shown as a star for a comparison. The arrows indicate the evolution directions of the models. }
\end{figure*}

\begin{figure}[htbp]
\centering
\includegraphics[scale=0.3]{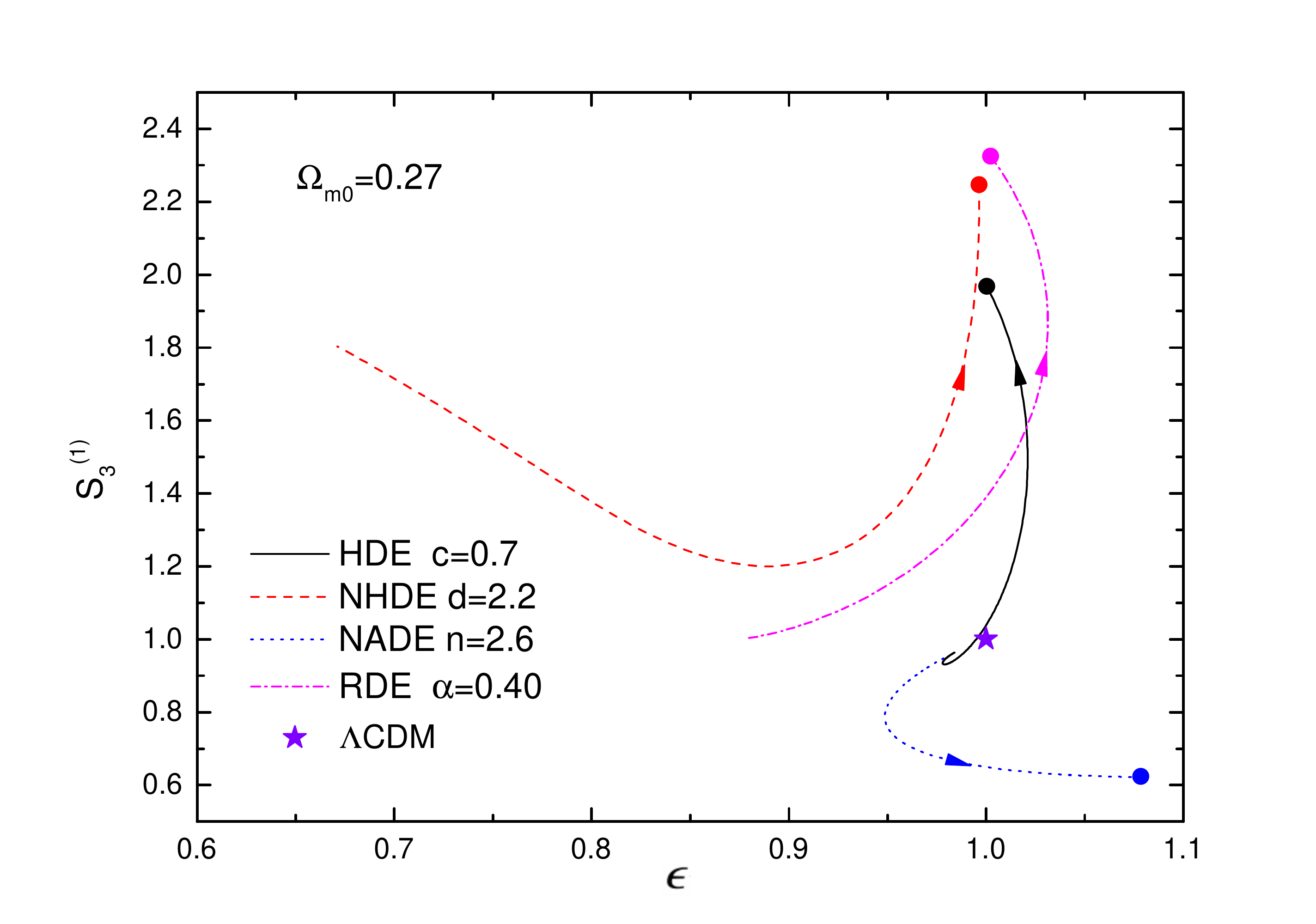}
\caption{\label{fig6} Comparisons of the evolutionary trajectories of $\{S^{(1)}_3, \epsilon\}$ of the HDE, NHDE, NADE and RDE models in the $S^{(1)}_3$--$\epsilon$ plane.
The current values of $\{S^{(1)}_3, \epsilon\}$ of the holographic DE models are marked by the round dots.
$\{S^{(1)}_3, \epsilon\}=\{1,1\}$ for the $\Lambda$CDM model is also shown as a star for comparison. The arrows indicate the evolution directions of the models. }
\end{figure}

\begin{figure*}[htbp]
\centering
\includegraphics[scale=0.5]{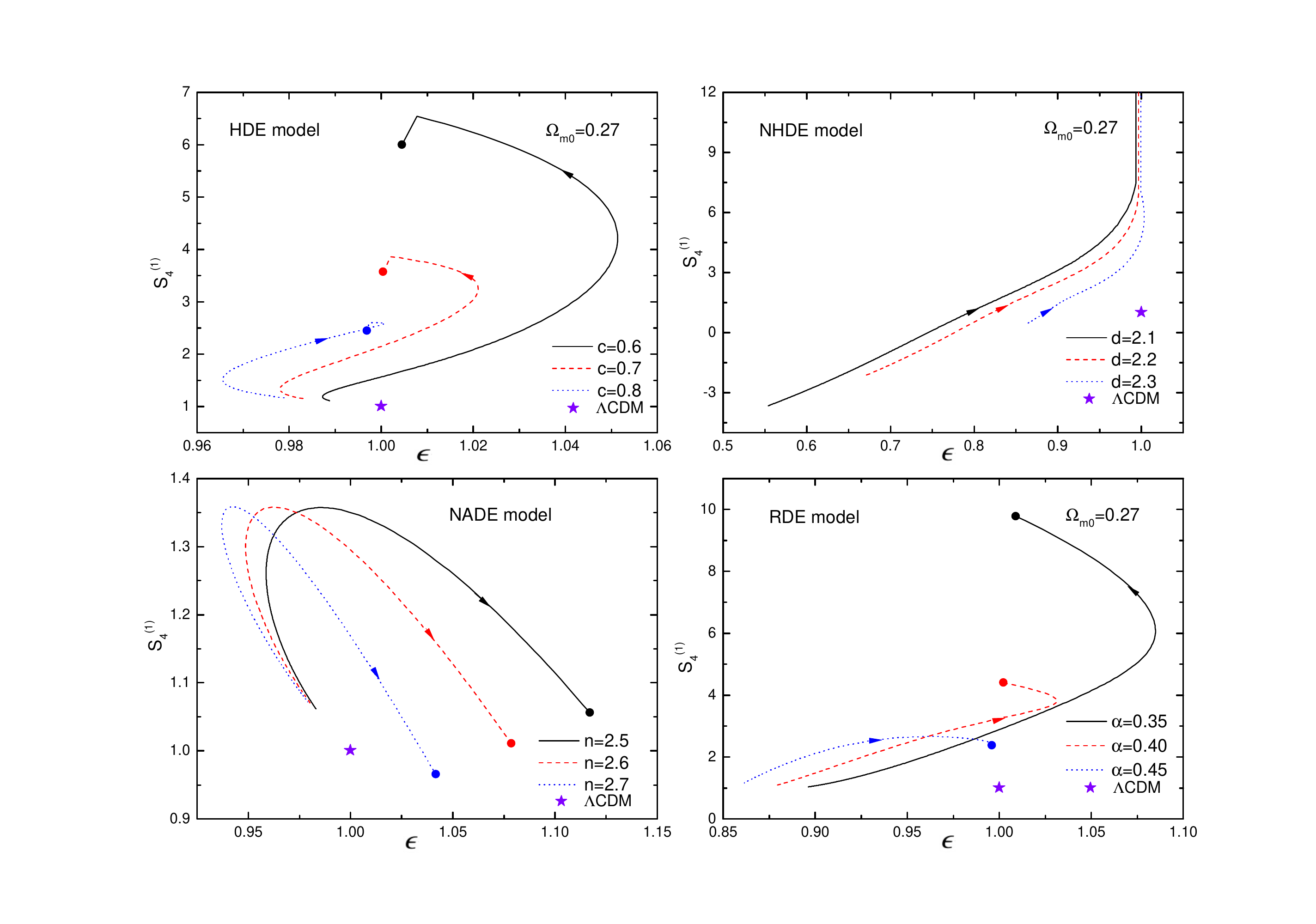}
\caption{\label{fig7} The composite null diagnostics $\{S^{(1)}_4, \epsilon\}$ are plotted for the HDE, NHDE, NADE and RDE models.
The current values of $\{S^{(1)}_4, \epsilon\}$ of the holographic DE models are marked by the round dots.
$\{S^{(1)}_4, \epsilon\}=\{1,1\}$ for the $\Lambda$CDM model is also shown as a star for a comparison. The arrows indicate the evolution directions of the models.
Note that the dots for current values of the NHDE model are not shown in this plot owing to the fact that the present-day $S_4^{(1)}$ values are too large compared to that of $\Lambda$CDM.}
\end{figure*}

\begin{figure}[htbp]
\centering
\includegraphics[scale=0.3]{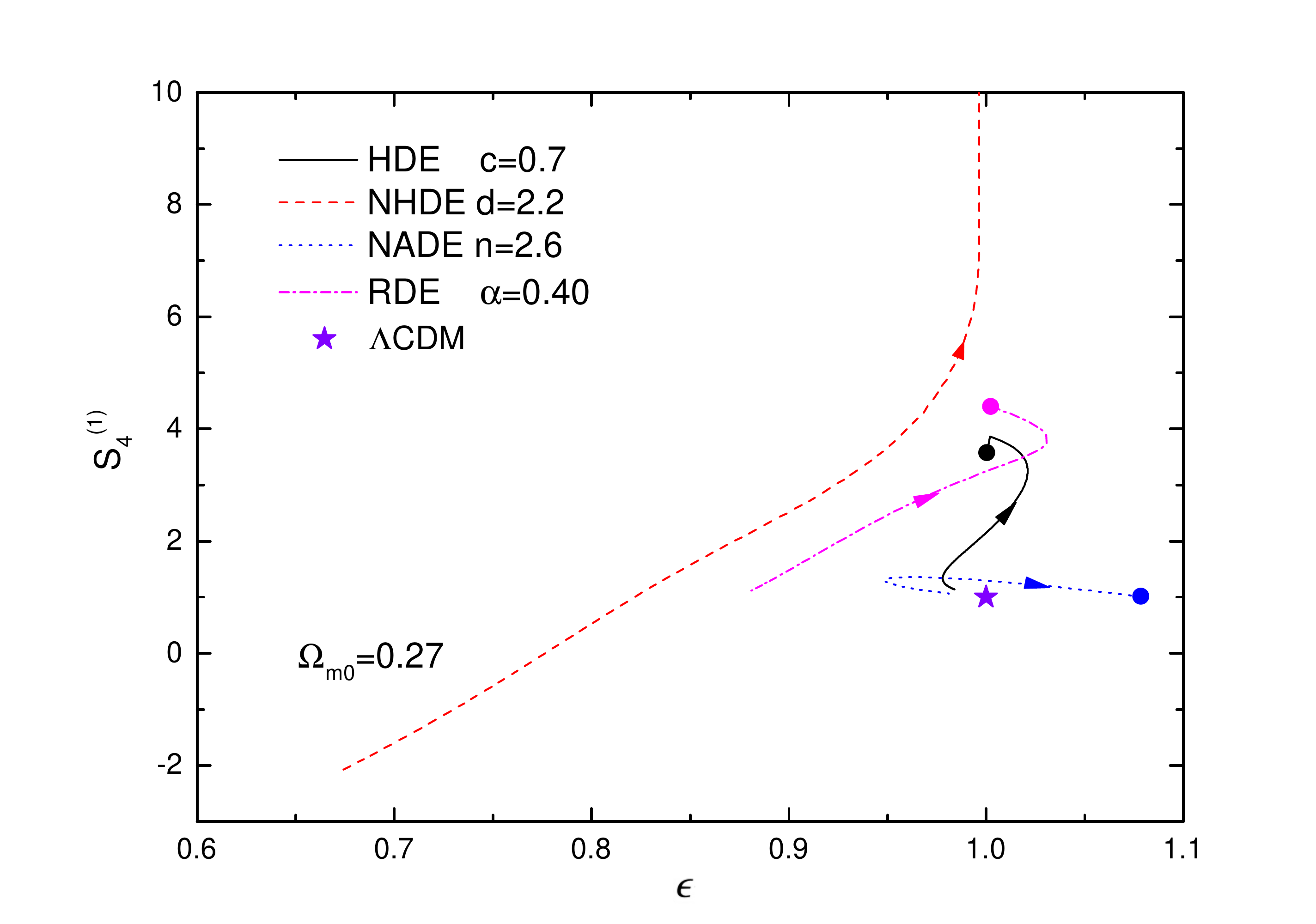}
\caption{\label{fig8} Comparisons of the evolutionary trajectories of $\{S^{(1)}_4, \epsilon\}$ of the HDE, NHDE, NADE and RDE models in the $S^{(1)}_4$--$\epsilon$ plane.
The current values of $\{S^{(1)}_4, \epsilon\}$ of the holographic DE models are marked by the round dots.
$\{S^{(1)}_4, \epsilon\}=\{1,1\}$ for the $\Lambda$CDM model is also shown as a star for comparison. The arrows indicate the evolution directions of the models.
Note that the dot for current values of the NHDE model is not shown in this plot owing to the fact that the present-day $S_4^{(1)}$ value is too large compared to that of $\Lambda$CDM.}
\end{figure}

\subsection*{2.2 The growth rate of perturbations}
The fractional growth parameter $\epsilon (z)$~\cite{Acquaviva:2008qp} can also be used as a null diagnostic, which is defined as
\begin{equation}
\epsilon(z)=\frac{f(z)}{f_{\Lambda CDM}(z)},
\end {equation}
where $f(z)=d\ln\delta/d\ln a$ describes the growth rate of the linear density perturbation~\cite{Wang:1998gt},
\begin{equation}
f(z)\simeq \Omega_m(z)^\gamma,
\end {equation}
\begin{equation}
\gamma(z)=\frac{3}{5-\frac{w}{1-w}}+\frac{3(1-w)(1-\frac{3}{2}w)}{125(1-\frac{6}{5}w)^3}(1-\Omega_m(z)),
\end {equation}
where $w$ either is constant, or varies slowly with time. For the $\Lambda$CDM model, $\gamma\simeq0.55$ and $\epsilon(z)=1$~\cite{Wang:1998gt,Linder:2005in}. However, for other models, the values of $\gamma$ and $\epsilon(z)$ depart from $\Lambda$CDM. For this reason, the fractional growth parameter $\epsilon (z)$ can be combined with the statefinders to define a composite null diagnostic (CND) $\{S_{\emph{n}}, \epsilon\}$~\cite{Arabsalmani:2011fz}. Obviously, we have $\{S_{\emph{n}}, \epsilon\} = \{1,1\}$ for $\Lambda$CDM.

\section {Holographic dark energy models}
Based on the holographic principle, the dark energy density is defined as $\rho _{de} = 3c^2M^2_pL^{-2}$~\cite{Li:2004rb,Cohen:1998zx}, where $c$ is an introduced numerical constant characterizing some uncertainties in the effective quantum field theory, and $L$ is the infrared (IR) cutoff in the theory. A series of DE models originating from the holographic principle were proposed. In this paper, we focus on the following models: HDE, NHDE, NADE, and RDE, and we describe them briefly in this section.

\subsection*{3.1 The HDE model}

In the HDE model~\cite{Li:2004rb}, $\rho _{de} = 3c^2M^2_pL^{-2}$, and $L$ is the future event horizon given by
\begin{equation}
L=a\int_{\it a}^\infty\frac{da'}{Ha'^2}.
\end{equation}
Here, the prime is used to differentiate the integration variable from the lower limit in the integral.
Note that throughout the paper, prime in integrals plays the same role as here.
In this model, $\Omega_{de}$ is described by the differential equation
\begin{equation}
\Omega '_{de}= \Omega_{de}(1- \Omega_{de})(1+\frac{2}{c}\sqrt{ \Omega_{de}}),
\end{equation}
where the prime denotes differentiation with respect to $\ln a$,
and $w$ is given by
\begin{equation}
w = -\frac{1}{3}-\frac{2}{3c}\sqrt{ \Omega_{de}}.
\end{equation}

\subsection*{3.2 The NHDE model}
In $2012$, the HNDE model in light of the action principle was proposed~\cite{Li:2012xf}, in which the dark energy density reads
\begin{equation}
\rho_{de}=M^2_p(\frac{d}{a^2L^2}+\frac{\lambda}{2a^4}),
\end{equation}
where $d$ is a numerical parameter, and
\begin{equation}
L=\int_{\it t}^\infty\frac{dt'}{a(t')}+L(a=\infty),~~~\dot{L}=-\frac{1}{a},
\end{equation}
\begin{equation}
\lambda=\int_0^t\frac{4a(t')ddt'}{L^3(t')}+\lambda(a=0),~~~\dot{\lambda}=-\frac{4ad}{L^3}.
\end{equation}
 For the NHDE model, $\Omega_{de}$ and $w$ can be given by
\begin{equation}
\Omega_{de}=-\frac{1}{3E^2}(\frac{da^{-2}}{\tilde{L}^2}+\frac{\tilde{\lambda}a^{-4}}{2}).
\end{equation}
\begin{equation}
w=\frac{\tilde{\lambda}\tilde{L}^2-2da^2}{3\tilde{\lambda}\tilde{L}^2+6da^2},
\end{equation}
where $\tilde{L}\equiv H_0L$, $\tilde{\lambda}\equiv \lambda/H^2_0$, and $E=H/H_0$.

\subsection*{3.3 The NADE model}
In the NADE model~\cite{Wei:2007ty},  $\rho _{de} = 3n^2M^2_p\eta ^{-2}$, where $n$ is a numerical parameter introduced, and the IR cutoff is provided by the conformal time $\eta$,
\begin{equation}
\eta =\int_0^a\frac{da'}{Ha'^2}.
\end{equation}
In this case, $ \Omega_{de}$ is the solution of the following differential equation
\begin{equation}
\Omega '_{de}= \Omega_{de}(1- \Omega_{de})(3-\frac{2}{na}\sqrt{ \Omega_{de}}),
\end{equation}
where the prime denotes differentiation with respect to $\ln a$,
and $w$ is given by
\begin{equation}
w = -1+\frac{2}{3na}\sqrt{ \Omega_{de}}.
\end{equation}

\subsection*{3.4 The RDE model}

In the RDE model~\cite{Gao:2007ep}, the IR cutoff $L$ is connected to Ricci scalar curvature,
 $\mathcal{R}=-6(\stackrel{\centerdot}{H}+2H^2)$.
 So Ricci dark energy density is
\begin{equation}
\rho_{de}=3\alpha M^2_p(\stackrel{\centerdot}{H}+2H^2),
\end{equation}
where $\alpha $ is a dimensionless coefficient.
Accordingly, one can get $\Omega_{de}$ and $w$ of RDE:
\begin{equation}
\Omega_{de}=\frac{1}{E^2}\frac{\rho_{de}}{\rho_0}=\frac{1}{E^2}\left(\frac{\alpha}{2-\alpha}\Omega_{m0}e^{-3x}+f_0e^{-(4-\frac{2}{\alpha})x}\right),
\end{equation}
\begin{equation}
w=\frac{\frac{\alpha-2}{3\alpha}f_0e^{-(4-\frac{2}{\alpha})x}}{\frac{\alpha}{2-\alpha}\Omega_{m0}e^{-3x}+f_0e^{-(4-\frac{2}{\alpha})x}},
\end{equation}
where $f_0=1-\frac{2}{2-\alpha}\Omega_{m0}$ is an integration constant.

\section {Statefinder hierarchy diagnostic}
For all models we fix $\Omega_{m0}=0.27$. To properly choose typical values of the parameters, we refer to the current observational constraints on the models. In HDE, the parameter $c$ takes 0.6, 0.7 and 0.8~\cite{Wang:2012uf}. In NHDE, $d$ takes 2.1, 2.2 and 2.3~\cite{Li:2012fj}. Since NADE is a single-parameter model, we apply the initial condition $\Omega_{de}(z_{ini})=n^2(1+z_{ini})^{-2}/4$ at $z_{ini}=2000$~\cite{Wei:2007xu}, and $n$ takes 2.5, 2.6 and 2.7~\cite{Zhang:2013lea}. In RDE, we choose $\alpha=0.35$, 0.40 and 0.45~\cite{Zhang:2009un}.

Firstly, the evolutions of $S^{(1)}_3$ versus redshift $z$ for the holographic DE models are plotted in Fig.~\ref{fig1}, and those of $\Lambda$CDM are also shown for comparison. We can see that in the low-redshift region the holographic DE models can easily be differentiated from the $\Lambda$CDM model, although in the high-redshift region they all but the NHDE model are nearly degenerate with the $\Lambda$CDM model. Furthermore, the difference between various values of parameter in one model can be directly identified for HDE and RDE in the low-redshift region, and for NHDE in the range of $z>0.5$. However, for NADE, the cases with different parameter values degenerate highly in both the low-redshift and the high-redshift region. Note that the $S^{(1)}_3$ diagnostic for the holographic DE models has been discussed in our previous work~\cite{Cui:2014sma}, and we repeat the relevant discussion in this paper for making the paper self-contained.

For breaking the degeneracy of NADE, in this paper we take into account $S^{(1)}_4$ from the statefinder hierarchy diagnostic~\cite{Arabsalmani:2011fz}, which includes the fourth-derivatives of $a(t)$. In Fig.~\ref{fig2}, the evolutions of $S^{(1)}_4$ versus redshift $z$ for the holographic DE models are plotted, and those of $\Lambda$CDM are also shown for comparison. From Fig.~\ref{fig2}, on one hand, the differences between the holographic DE models and the $\Lambda$CDM model become clearer in the low-redshift region, although the curves of HDE and RDE degenerate with those of $\Lambda$CDM in the high-redshift region but which is slighter than that of Fig.~\ref{fig1}. On the other hand, it is important to see that the degeneracy in the NADE model with different parameter values appearing in Fig.~\ref{fig1} is broken, and for HDE, NHDE, and RDE models the cases with different parameter values can be discriminated more evidently in comparison with those of Fig.~\ref{fig1}.

The same conclusion can also be drawn from Table~\ref{tab1}, in which we show the today's values of statefinders, $S^{(1)}_{3to}$ and $S^{(1)}_{4to}$, and the differences of them, $\Delta S^{(1)}_{3to}$ and $\Delta S^{(1)}_{4to}$, for the holographic DE models, where $\Delta S^{(1)}_{3to}=S^{(1)}_{3to}(\rm{max})-S^{(1)}_{3to}(\rm{min})$ and $\Delta S^{(1)}_{4to}=S^{(1)}_{4to}(\rm{max})-S^{(1)}_{4to}(\rm{min})$ within one model. The current values of the statefinders also play an important role in diagnosing different DE models. In Table~\ref{tab1}, we can see that the differences between different parameter values in one model are magnified through $S^{(1)}_{4to}$, because the values of $\Delta S^{(1)}_{4to}$ are remarkably bigger than those of $\Delta S^{(1)}_{3to}$ for most cases. For the NADE model, $S^{(1)}_{4to}=0.09$, only slightly larger than $\Delta S^{(1)}_{3to}=0$, which also indicates a weak degeneracy for different parameter values.

For further comparing the statefinders $S^{(1)}_3$ and $S^{(1)}_4$, we make comparisons of the holographic DE models and the $\Lambda$CDM model in the $S^{(1)}_3$($z$) plots (Fig.~\ref{fig3}) and in the $S^{(1)}_4$($z$) plots (Fig.~\ref{fig4}). From these two figures, we find that the differentiable redshift region of the various DE models extends from $z\sim 0$ -- 1 in the $S^{(1)}_3$($z$) plots to $z\sim 0$ -- 3 in the $S^{(1)}_4$($z$) plots, and $S^{(1)}_4$ leads to more apparent distinctions for the various DE models. In addition, from Table~\ref{tab1}, the current values of $S^{(1)}_{4}$ for different holographic DE models separate more distinctively than those of $S^{(1)}_{3}$. Therefore, $S^{(1)}_4$ can diagnose different holographic DE models more effectively.

The above analysis shows that there still is a weak degeneracy for the NADE model with different parameter values even when the $S^{(1)}_4$ diagnostic is employed. So we consider the conjunction of the geometrical diagnostic (the statefinder hierarchy) and the cosmic growth history diagnostic (i.e., the fractional growth parameter $\epsilon (z)$), instead of only using the geometrical diagnostic. Acting as an alternative null diagnostic, the CND $\{S_{\emph{n}}, \epsilon\}$~\cite{Arabsalmani:2011fz} is studied in this paper.

The evolutionary trajectories of $\{S^{(1)}_3, \epsilon\}$ for the holographic DE models are plotted in Fig.~\ref{fig5}, where the present-day values of $\{S_3^{(1)},\epsilon\}$
for the models are marked by the round dots, and the fixed point $\{S^{(1)}_3, \epsilon\}=\{1,1\}$ for the $\Lambda$CDM model is also shown as a star for comparison.
The arrows indicate the evolution directions of the models.
The difference between the specific holographic DE model and the $\Lambda$CDM model is measured by the separation of the round dot and the star, and
the differences of the cases in one model with different parameter values can also be measured by the separations between the dots.
We find that, by employing the CND, $\{S^{(1)}_3, \epsilon\}$, the differences between the evolving curves of the
various holographic DE models and the fixed point of $\Lambda$CDM are fairly evident. For the HDE and the RDE models, their $\epsilon_0$ values are all around 1,
but their present-day $S^{(1)}_3$ values are capable of differentiating their cases with different parameter values and them from $\Lambda$CDM.
On the contrary, for the NADE model, the present-day $S^{(1)}_3$ values are in degeneracy, but its $\epsilon_0$ values are distinctively different,
so the degeneracy of NADE is effectively broken by $\epsilon_0$ using the CND $\{S^{(1)}_3, \epsilon\}$.
For the NHDE model, since both the $S^{(1)}_{\rm3to}$ and the $\epsilon_0$ values are in degeneracy, discriminating the cases with different parameter values can only depend on evolutionary trajectories of $\{S_3^{(1)},\epsilon\}$.
For a direct comparison, we also plot the evolutionary trajectories for the various holographic DE models and the fixed point for the $\Lambda$CDM model
in the $S^{(1)}_3$--$\epsilon$ plane in Fig.~\ref{fig6}.

Furthermore, we also apply the CND $\{S^{(1)}_4, \epsilon\}$ to study the holographic DE models.
We plot the evolutionary trajectories for the holographic DE models in the $S^{(1)}_4$--$\epsilon$ plane in Fig.~\ref{fig7}, where the present-day values of $\{S_4^{(1)},\epsilon\}$
for the models and the fixed point $\{1,1\}$ for $\Lambda$CDM are also shown as round dots and star, respectively, for directly measuring the ``effective distances'' between them.
Note that the dots for the current values of the NHDE model are not shown in this plot owing to the fact that the present-day values of $S^{(1)}_4$ are too large
compared to that of $\Lambda$CDM (from Table~\ref{tab1}, one can see that $S^{(1)}_{4to}\simeq 160$--190 for NHDE). 
For the NHDE model, due to the fact that $\Delta S^{(1)}_{4to}\gg\Delta S^{(1)}_{3to}$ (see Table \ref{tab1}), the $S^{(1)}_{4to}$ values in this CND are used to discriminate the cases of NHDE with 
different parameter values. 
From this figure, one can also clearly see that, using the combination of the statefinder hierarchy $S^{(1)}_4$ and the fractional growth parameter $\epsilon$,
the degeneracy of NADE can be further broken, since both $\Delta S^{(1)}_{4to}$ and $\Delta\epsilon_0$ are considerable (see also Table \ref{tab1}). 
Therefore, employing the CND $\{S^{(1)}_4, \epsilon\}$, all the holographic DE models under consideration
can be differentiated quite well. To make a clearer comparison of them, we show in Fig.~\ref{fig8} the evolutionary trajectories for the various holographic DE models and
the fixed point for the $\Lambda$CDM model in the $S^{(1)}_4$--$\epsilon$ plane.

\section {Conclusion}
In this paper, we diagnose the holographic DE models with the statefinder hierarchy that is essentially a series of null diagnostics. By using $S^{(1)}_4$ which contains fourth derivatives of $a(t)$, the holographic DE models are distinguished more evidently from one anther and from the $\Lambda$CDM model, compared to the results by using $S^{(1)}_3$. The analysis of the current values of $S^{(1)}_4$ also indicates that the statefinder $S^{(1)}_4$ performs better than $S^{(1)}_3$ for diagnosing the holographic DE models. We also consider the CND, $\{S^{(1)}_3, \epsilon\}$, combining the statefinder  $S^{(1)}_3$ with the fractional growth parameter $\epsilon (z)$, and find that the CND $\{S^{(1)}_3, \epsilon\}$ is a rather useful diagnostic approach. 
Furthermore, we apply the CND, $\{S^{(1)}_4, \epsilon\}$, combining the statefinder hierarchy $S^{(1)}_4$ with $\epsilon$, to study the holographic DE models, and we find that 
$\{S^{(1)}_4, \epsilon\}$ is even much better than $\{S^{(1)}_3, \epsilon\}$ in discriminating the different cases within one model with different parameter values.
Our results demonstrate that the statefinder hierarchy containing higher derivatives of $a(t)$ and the CND are fairly useful in distinguishing the holographic DE models from one another as well as from the $\Lambda$CDM model, and the CND is highly efficient for breaking the degeneracies for different parameter values in one model.

\section*{Acknowledgements}
We wish to thank the referee for providing us with valuable comments and suggestions, helping us improve the work significantly. 
This work was supported by the Provincial Department of Education of Liaoning under Grant No. L2012087, the National Natural Science Foundation of China under Grant No. 11175042 and the Fundamental Research Funds for the Central Universities under Grant No. N120505003.

\end{document}